\pdfoutput=1
\documentclass[10pt, twocolumn, aps, prb, superscriptaddress, floatfix, showpacs, notitlepage]{revtex4-2}

\usepackage[utf8]{inputenc}
\usepackage[T1]{fontenc}
\usepackage{lmodern}
\usepackage{graphicx}
\usepackage{amsmath}
\usepackage{amssymb}
\usepackage{extarrows}
\usepackage{bm}
\usepackage{xcolor}
\usepackage[colorlinks, citecolor={blue!50!black}, urlcolor={blue!50!black}]{hyperref}
\usepackage{bookmark}
\usepackage{tabularx}
\usepackage{mathtools}
\usepackage{microtype}
\usepackage{siunitx}
\usepackage{enumitem}
\usepackage{booktabs}
\usepackage{MnSymbol} 

\newcommand{\kk}{{\bm{\kappa}}}
\newcommand{\kp}{{\bm{\kappa '}}}
\newcommand{\vv}{\bm{v}}
\newcommand{\vE}{\bm{E}}
\newcommand{\vk}{{\bm{v}_{\bm{\kappa}}}}
\newcommand{\vx}{\bm{r}}
\newcommand{\LL}{\bm{\Lambda}}
\newcommand{\rp}{\bm{\rho}}
\newcommand{\pd}{{\phantom{\dagger}}}

\graphicspath{{figures/}}

\begin{document}

\author{Pablo M. Perez-Piskunow}
\affiliation
{Catalan Institute of Nanoscience and Nanotechnology (ICN2),
CSIC and BIST, Campus UAB, Bellaterra, 08193 Barcelona, Spain}
\author{Nicandro Bovenzi}
\affiliation
{Instituut-Lorentz, Universiteit Leiden, P.O. Box 9506, 2300 RA Leiden, The Netherlands}
\author{Anton R. Akhmerov}
\affiliation
{Kavli Institute of Nanoscience, Delft University of Technology,
P.O. Box 4056, 2600 GA Delft, The Netherlands}
\author{Maxim Breitkreiz}
\email{breitkr@physik.fu-berlin.de}
\affiliation{Dahlem Center for Complex Quantum Systems and Fachbereich Physik, Freie Universit\" at Berlin, 14195 Berlin, Germany}

\title{
 Chiral Anomaly Trapped in Weyl Metals: Nonequilibrium Valley Polarization
\texorpdfstring{\\}{}
at Zero Magnetic Field
}

\begin{abstract}

In Weyl semimetals the application of parallel electric and magnetic
fields leads to valley polarization---an
occupation disbalance of valleys of opposite
chirality---a direct consequence of the chiral anomaly.
 In this work, we present numerical tools to 
explore such nonequilibrium effects
in spatially confined three-dimensional 
systems with a variable 
disorder potential, giving exact solutions to leading 
order in the disorder potential and the applied electric field. 
Application to a Weyl-metal slab  shows that valley polarization
also occurs without an external magnetic field as
  an effect of 
 chiral anomaly ``trapping'': Spatial confinement produces 
 chiral bulk states, which enable the 
 valley polarization in a similar way as the chiral 
 states induced by a magnetic field.
Despite its finite-size origin,
the valley polarization can persist up to macroscopic
length scales if the disorder potential is sufficiently long ranged,
so that  direct inter-valley scattering is  suppressed and
the relaxation then goes via the Fermi-arc surface states.

\end{abstract}

\maketitle

\section{Introduction}
The most famous effect associated with Weyl Fermions is the chiral
anomaly~\cite{Adler1969, Bell1969}---magnetic-field induced chiral states moving
parallel or antiparallel to the field, depending on the chirality of the Weyl
Fermion.
In Weyl semimetals~\cite{Armitage2017, Yan2017} the two chiralities occur
pairwise, ensuring an equal number of forward- and backward-propagating states,
and the chiralities   are connected by Fermi-arc surface states.
Existing Weyl-semimetal materials  typically have a small but finite Fermi momentum
$k_F$ measured from the Weyl node and a much larger 
 momentum-space separation
$\Delta k$  of valleys that host the opposite chiralities.

The valley degree of freedom  \cite{Rycerz2007}
 plays a central role 
in the transport behavior of Weyl semimetals. 
Parallel electric and magnetic fields  produce a difference
in the non-equilibrium occupation of 
valleys~\cite{Parameswaran2014, Armitage2017, Bednik2020}.
A direct consequence of this valley polarization is an enhanced
conductivity parallel to the magnetic field due to the
polarization-enhanced occupation disbalance of countermoving chiral
states
~\cite{Nielsen1983,Fukushima2008,Son2013,Burkov2014a,Spivak2016,Behrends2017}.
Experimental observations, although obscured by the competing current-jetting
effect~\cite{Reis2016}, support the general feature of a chiral-anomaly
enhancement of the conductivity~\cite{Xiong2015, Hirschberger2016, Liang2018}.
Other manifestations of the valley degree of freedom are found in nonlocal transport 
measurements  \cite{Parameswaran2014, Zhang2017} and in the photogalvanic response \cite{Ma2017,Chan2017a}. 

Crucial in understanding ``valleytronic'' transport  is to 
explore the effect of disorder and the finite size of the crystal, 
which are two 
unaviodable properties of real materials.   
Disorder plays a subtle role if the Fermi level lies at the Weyl nodes, where 
it may or may not destroy the ideal semimetal
phase  by inducing a finite density of states 
\cite{Sbierski2014, Buchhold2018, Syzranov2018, Wilson2020} or, for 
finite inter-valley scattering, drive the system into an insulating phase  \cite{Chen2015a}.
At finite chemical potentials, well-separated Weyl nodes, and a weak 
disorder potential the Weyl-semimetal phase has proven to be robust, allowing 
for a perturbative treatment of disorder, which will be employed in this work.

\begin{figure}[b]
\includegraphics[width=\linewidth]{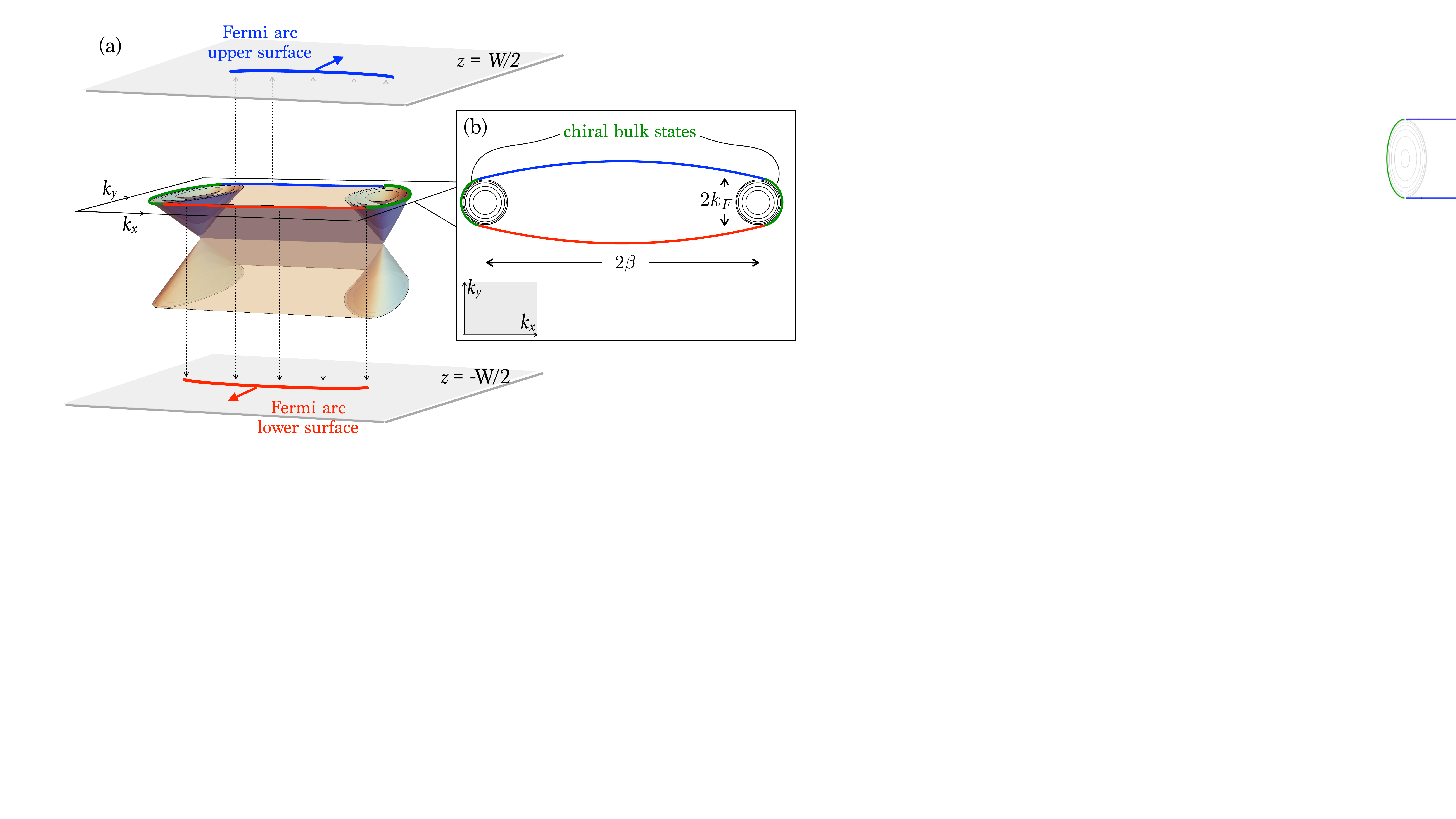}
\caption{
(a) Schematic picture of the considered Weyl-slab model
in a mixed momentum-/real-space illustration.
The plot shows the slab
spectrum as a function of in-plane momenta.
Top (blue) and bottom (red) surface states are indicated with their position and propagation
direction in real space.
Chiral bulk states connecting surface states of
both surfaces are depicted
in green. (b) Top-view on equi-energy contours of the
slab spectrum of (a).}
\label{setup}
\end{figure}

Finite-size effects in meso- and macroscopic
Weyl semimetals (crystal dimensions much larger than the lattice constant),
have also been explored with the focus 
on the role of topological  Fermi-arc surface states
 \cite{Gorbar2016a, Igarashi2017, Resta2018,
 Bovenzi2017a, Gorbar2018, Sukhachov2019,
 Kaladzhyan2019,  Behrends2019, Breitkreiz2019,
 Zhang2019, Breitkreiz2020}.  Peculiarities
are rooted in the specifics of the momentum-space structure of Fermi arcs
 connecting valleys of opposite chirality and their 
 unidirectional motion at a single surface, see Fig.~\ref{setup}.
Separation of countermoving Fermi arcs to 
opposite surfaces explains their relevance at large systems sizes, most 
prominently in the intrinsic anomalous Hall effect
 \cite{Burkov2014, Suzuki2016, Li2019, Breitkreiz2019}.  
The relevance of finite-size effects for the valley degree of freedom,
 on the other hand,  is much less obvious, since the valleys 
 consist of extended bulk states, lacking the spatial separation. 

In this work, we show that in a disordered Weyl semimetal slab 
valley polarization can be induced without external magnetic fields
as a finite-size effect at mesoscopic slab widths, possibly extending
 to even larger sizes. Crucial turn out to be 
 confinement-induced chiral bulk states
 \cite{Behrends2019, Breitkreiz2020}:
  At zero magnetic field and a finite Fermi momentum $k_F$
there is a residual density of chiral bulk states,
which must remain to reconnect the two Fermi-arc surface states
as shown in Fig.~\ref{setup}.
The density of  chiral bulk states of a single valley,
relative to the density of magnetic-field induced chiral
states, is $k_F / l_B^{-2}W$, where
$l_B = \sqrt{\hbar/eB} \approx 26\, \textrm{nm} \sqrt{1/B[\mathrm{T}]}$
is the magnetic length, and $W$ the width of the slab.
Taking an experimentally realistic value of $k_F=\SI{0.01}{\angstrom^{-1}}$, the density of
anomalous chiral states is larger than that of magnetic-field induced ones
for $W \lesssim 100\,\textrm{nm}$ $/B[\mathrm{T}]$.
At a mesoscopic width $W \sim 100\,\textrm{nm}$, the effect of anomalous chiral
states is thus comparable to that of the field-induced chiral states at
$B \sim 1\,\mathrm{T}$,
which relevance is commonly accepted and experimentally supported
~\cite{Xiong2015, Hirschberger2016, Liang2018}.

We find that
the confinement-induced valley polarization and the presence of surface states
can lead to conductivity enhancements by several orders
of magnitude, compared to that of the infinite system.
This conductivity enhancement is suppressed
with increasing width $W$ as $1/W$, simply due to
the decreasing density of confinement-induced
chiral states. The valley polarization, however,  turns out to remain unsupressed up to widths set by
the probability of direct inter-valley scattering,
which in case of Gaussian-type disorder potetials is
exponentially enhanced $\sim \exp\big[(\xi\Delta k)^2\big]$.

 To reveal this effect, we develop a  two-part numerical approach combining
the full quantum mechanical calculations of a multiband slab dispersion and
wavefunctions,
with a numerical solution for the non-equilibrium corrections 
to the density matrix   \cite{zenodo}. 
 The resulting non-equilibrium density matrix is 
exact to leading order in the disorder potential 
and the applied external electric field. 

The paper is organized as follows. In Section~II we start with a derivation of
transport equations for multiband systems with a large number of bands and discuss its validity regime.
In Section~III we introduce the model of the Weyl  slab, the impurity potential,
and calculate the scattering rate.
In Section~IV we present the transport results obtained
from solving the transport equations numerically, which we discuss in Section~V by comparing with simplified
analytic calculations.
We conclude in Section~VI.

\section{Quantum transport approach}

In the first part of this section we recapitulate the general
transport formalism in the presence of weak disorder
following Kohn and Luttinger \cite{Kohn1957}. This is necessary
to identify the validity regime of this formalism when applied to a spatially confined system, which we do in the second part.

\subsection{General quantum transport approach}

We separate a general single-particle Hamiltonian into the free-particle
part $H_0$,
the additional weak
scattering potential $V$, and
a time-dependent electric-field term $e\vE\cdot \bm{r}\, e^{st}$ with the position
operator $\bm{r}$ and an adiabatic time dependence
$e^{st}$ with $s\to 0^+$,
\begin{equation}
H = H_0+V+ e^{st} e\vE\cdot \bm{r}.
\end{equation}
The scattering is due to a random configuration of impurities
with  a vanishing
impurity-averaged potential $\llangle  V \rrangle = 0$.

We make the ansatz for the full density matrix
\begin{equation}
    \rho = p+ g e^{st},
\end{equation}
where $p$ is the equilibrium density matrix
\begin{equation}
    p = \frac{e^{-\beta(H_0+V)}}{\mathrm{Tr}\,  e^{-\beta(H_0+V)}}
\end{equation}
and $g$ the non-equilibrium correction.
Inserting into the von Neumann equation for the density matrix,
\begin{equation}
    i \partial_t \rho = [H,\rho],
\end{equation}
and expanding to first order in $E$ we obtain
\begin{equation}
    i\,s\,g=[e\vE\cdot \bm{r},p]+ [H_0+V,g].
    \label{KL}
\end{equation}

The following analysis consists in expanding~\eqref{KL} in powers of $V$.
We write~\eqref{KL} in terms of its matrix elements
in the basis of $H_0$ eigenstates $|\kk \rangle$, where $\kk$ combines
the quantum numbers.  Off-diagonal and diagonal elements read, respectively,
\begin{gather}
    \begin{split}
        (E_\kk-E_{\kp} - is)\, g_{\kk\kp} =
        (g_\kk-g_\kp) V_{\kk\kp}+C_{\kk\kp} \\
        + \sum_{\kk''\neq \kp,\kk}\left(g_{\kk\kk''}V_{\kk''\kk'}-V_{\kk\kk''}g_{\kk''\kk'}\right),
    \end{split} \label{first} \\
    -is g_\kk = C_\kk +\sum_{\kp\neq \kk}(g_{\kk\kp}V_{\kp\kk}-V_{\kk\kp}g_{\kp\kk}),
    \label{second}
\end{gather}
where the notation is $A_{\kk\kp} = \langle \kk | A | \kp \rangle$,
$A_\kk = A_{\kk\kk}$, $E_\kk =  \langle \kk | H_0 | \kk \rangle $.
The field-dependent term
\begin{align}
    C_{\kk\kp} &= e \vE\cdot \left[\bm{r} , p\right]_{\kk\kp}
    \label{C}
\end{align}
expands in powers of $V$ starting with the  zeroth order,
\begin{equation}
    C^{(0)}_{\kk\kp}= e\bm{E}\cdot \bm{r}_{\kk\kp}\left[n_F(E_\kk)-
    n_F(E_{\kp})\right],
\label{C0}
\end{equation}
where $n_F(E)$ is the Fermi distribution.
From~\eqref{first} and~\eqref{second} we see that the leading order of the
off-diagonals of $g$ are of order $V^{-1}$, while the diagonals are of order
$V^{-2}$.
To leading order, the latter two terms in~\eqref{first}
can thus be neglected, leading to
\begin{equation}
    g_{\kk\kp} =\frac{ (g_\kk-g_\kp) V_{\kk\kp}}{E_\kk-E_{\kp}-is}.
\end{equation}
Inserting into~\eqref{second}, taking the adiabatic limit
$s\to 0^+$, and applying disorder averaging $\llangle \dots \rrangle$  we obtain
\begin{equation}
    C^{(0)}_\kk = i \, 2 \pi \sum_{\kp'\neq \kk}\delta(E_\kk-E_{\kp})
    \left\llangle|V_{\kk\kp}|^2\right\rrangle (g_{\kp}-g_{\kk}).
    \label{BE0}
\end{equation}

If the electric field points in a direction in which
the system is infinite, let it be $x$ and $y$,
 the eigenstates can be chosen as momentum
$\bm{k} = (k_x, k_y)$ eigenstates. The field term~\eqref{C0} becomes
\begin{equation}
    C^{(0)}_\kk= i\, e\bm{E}\cdot \vv_\kk \, n_F'(E_\kk),
\end{equation}
where $ \vv_\kk = \partial_\kk E_\kk$ is the velocity.

Making the ansatz
\begin{equation}
    g_\kk =  -e\bm{E}\cdot \LL_\kk \, n_F'(E_\kk)
    \label{ansa}
\end{equation}
\eqref{BE0} simplifies to
\begin{multline}
    n_F'(E_\kk) \vv_\kk  = 2 \pi \sum_{\kp\neq \kk}n_F'(E_\kk) \delta(E_\kk-E_{\kp})
    \\
    \times
    \left\llangle |V_{\kk\kp}|^2 \right\rrangle(\LL_{\kk}-\LL_{\kp}),
    \label{BE1}
\end{multline}
known as  Boltzmann equation (BE), to be solved with respect to the vector-valued
state-resolved transport mean free paths $\LL_\kk$,
which we will refer to as \emph{transport length} in short.
The average magnitude of the transport length scales with the
strength of the impurity potential as $   V^{-2}$.

Note that since the summation
operator acting on $\LL_\kk$ in~\eqref{BE1} has
an eigenvalue zero for a $\kk$ independent vector,
the solution is generally determined up to a constant
\begin{equation}
    \bm{c}=\frac{ \sum_\kk n_F'(E_\kk)\LL_\kk}{\sum_\kk n_F'(E_\kk)}.
\end{equation}
Particle conservation however requires $ \sum_\kk g_\kk = 0$, which fixes
the constant to $\bm{c}=0$.

The current-density expectation value reads
\begin{equation}
    \bm{j} = -e\frac{1}{V}\sum_\kk \bm{v}_\kk \, g_\kk,
\end{equation}
where $V$ is the system volume. The conductivity tensor $\sigma$, defined as
$\bm{j} = \sigma \bm{E}$, becomes, using~\eqref{ansa},
\begin{equation}
    \sigma = e^2\frac{1}{V}\sum_\kk n_F'(E_\kk) \bm{v}_\kk  \otimes \LL_{\kk}.
    \label{sigma}
\end{equation}
Note that the BE~\eqref{BE1} is
exact in the weak-disorder limit, giving a conductivity that scales
with the squared inverse strength of the disorder potential. Leading corrections,
which will not be considered here, are of zeroth order in the
impurity potential, they include, e.g., the anomalous Hall effect.

\subsection{Application to a slab model}

We now discuss the validity regime of~\eqref{BE1} when applied to a
slab model. We consider a system that is infinite in
two spatial directions, $x$ and $y$ (as specified above), and confined in direction $z$ to $-W/2 < z < W/2$.
The slab energy eigenspace
$\kk  =  (\bm{k},b)$, where $\bm{k}$
is the in-plane momentum, has the
particularity that the number of bands (band index $b$)
is potentially very large, scaling with
the width $W$ of the system.  Since the BE that we
have just derived
relies only on considering the leading order in the scattering
potential $V$, it can still be applied to the slab, provided that
$V$ can be taken to be
arbitrary small. For the slab, a problem arises if we want
to consider such a large width $W$ that  the effect of
boundaries becomes smaller than that of the impurity scattering,
which can invalidate the expansion in powers of $V$.
We now examine when exactly the width
becomes ``too large'' in that sense.

The large
width $W$ enters our above formalism through the position matrix elements in
$C_{\kk\kp}$, see Eq.~\eqref{C}. Let us thus repeat the above steps without
neglecting higher $V$ orders in  $C$ since they might still be
large due to $W$. In this case the BE obtains an extra term on the
right-hand side (rhs), so that the extended BE reads
\begin{multline}
    C_\kk = \sum_{\kk'\neq \kk}\delta(E_\kk-E_{\kp}) |V_{\kk\kp}|^2 (g_{\kp}-g_{\kk})\\
    +\sum_{\kp\neq \kk}\delta(E_\kk-E_{\kp}) (C_{\kk\kp}V_{\kp\kk}-V_{\kk\kp}C_{\kp\kk}).
    \label{BEex}
\end{multline}
Now expanding $C$ in powers of $V$ the rhs term with $C^{(0)}$ vanishes upon
impurity averaging since the mean potential due to impurities is zero.
We thus consider the next order term,
\begin{multline*}
    C^{(1)}_{\kk\kp} =
    e\vE\cdot \sum_{\kk''} \left[\bm{r}_{\kk\kk''} V_{\kk''\kk'}
    \frac{n_F(E_{\kk''})  -n_F(E_{\kk'})}{E_{\kk''} - E_{\kk'}}\right.
    \\
    \left. -
    \bm{r}_{\kk''\kk'} V_{\kk\kk''}
    \frac{n_F(E_{\kk})  -n_F(E_{\kk''})}{E_{\kk} - E_{\kk''}}\right].
\end{multline*}
While it also vanishes upon averaging on the left-hand side (lhs) of
\eqref{BEex}, the new term on  the rhs becomes
\begin{multline*}
e\vE\cdot \sum_{\kk'\neq \kk}\sum_{\kk''} \delta(E_\kk-E_{\kk'})
\frac{n_F(E_{\kk})  -n_F(E_{\kk''})}{E_{\kk} - E_{\kk''}} \\
\times \left[
V_{\kk''\kk'}V_{\kk'\kk}\bm{r}_{\kk\kk''}+V_{\kk\kk'}V_{\kk\kk''}\bm{r}_{\kk''\kk'}\right].
\end{multline*}
In this sum there are terms that are proportional to
$|V_{\kk\kk'}|^2$, which certainly do not vanish upon
impurity averaging. Compared to the first term on the rhs of~\eqref{BEex}
with the general ansatz~\eqref{ansa},
the new term is generally smaller if the position matrix elements
are smaller than the typical values of the transport length,
let us denote them by $\bar{\Lambda}$.
This correction can thus be neglected only if the width is much smaller,
\begin{equation}
W \ll \bar{\Lambda}.
\label{wl}
\end{equation}
Higher order terms due to the expansion of $C$ are of order $W$ times a higher
power of $V$ and thus give even smaller corrections.

Summarizing section III, Eq.~\eqref{wl}
characterizes the validity regime of the BE~\eqref{BE1} if applied to the slab.
In words, one is allowed to consider impurity scattering as a weak perturbation
to a free propagation in the slab treated as a two-dimensional multiband system
as long as the mean free path is much larger than the width.

\section{Weyl-semimetal slab model}

\begin{figure}
\includegraphics[width=\linewidth]{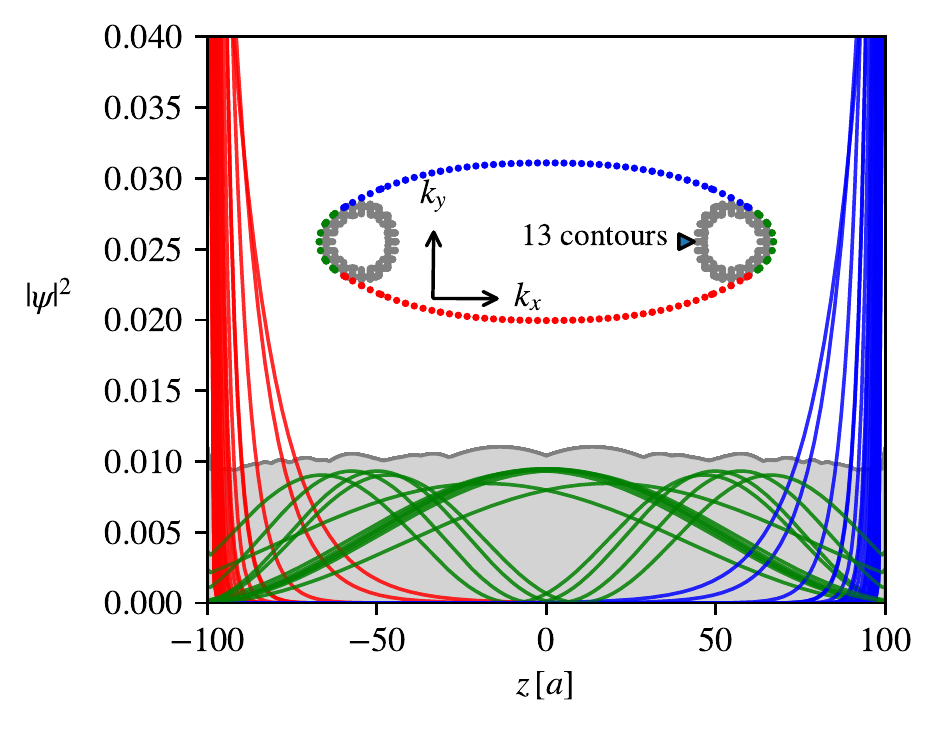}
\caption{
Position-$z$-resolved probability density $|\psi|^2= |\langle z|\psi\rangle|^2$
of Fermi-level states.
Colors indicate surface states at the lower boundary (red) and upper boundary (blue),
the anomalous chiral bulk states (green), and normal bulk states (gray).
The  inset shows the position of the states in momentum space.
Other parameters are $E_F=0.3\, t,\; W=201,\; \mu_b=0.3\, t$. }
\label{fig:waves}
\end{figure}

We consider a minimal lattice model of a Weyl semimetal~\cite{Yang2011},
\begin{align}
    H_0(\kk) &= t\sigma_x\sin{k_z} + t\sigma_y\sin{k_y}
    + m_\kk\sigma_z,
    \label{ham0}
\end{align}
where $m_\kk = t (2+\cos{\beta} - \cos{k_x} -\cos{k_y} - \cos{k_z})$ and
$\sigma_i$ are pseudospin Pauli matrices (corresponding to
an arbitrary degree of freedom), $t$ is the hopping amplitude, and the lattice
constant is set to unity.
The two Weyl nodes are placed at $\bm{k}=\pm \bm{ \beta}$, where
$\bm{\beta} = \beta \hat{\bm{x}}$ corresponds to a time-reversal breaking magnetization. We consider a ``good''
Weyl semimetal with a cone separation $\beta\sim 1$.

The Hamiltonian of the slab is given by the lattice Hamiltonian~\eqref{ham0}
but for a finite number $W$ of sites in the $z$ direction. Transformation into
the site basis in the
$z$ direction replaces $\cos k_z \to (\delta_{i,j+1}+\delta_{i,j-1})/2$,
where $i =0,1,\dots, (W-1)$ is the site number, corresponding to the
discrete position in $z$,
\begin{equation}
z \equiv i-\frac{W-1}{2},
\end{equation}
in units of the lattice constant which is set to one. We furthermore
add a boundary potential $\mu_b$ at the surface layers of the slab,
which main effect is to bend the Fermi-arc surface states.
We label the eigenstates by $\kk = (\bm{k},b)$ where $\bm{k}=(k_x,k_y)$ are the
continuous in-plane momenta and $b$ denotes the $2W$ modes at each value of
$\bm{k}$.

The eigenstates $ |\psi_\kk\rangle$ and eigenenergies $E_\kk$
of the slab are obtained from exact diagonalization of the Hamiltonian
at a fixed in-plane momentum $\bm{k}$ using standard
methods of numerical diagonalization \cite{zenodo}.
For our transport considerations we need to take into account all
Fermi-level states, which are continuous contours in the space of
the in-plane momentum $\bm{k}$.
We numerically \cite{zenodo} determine the
 contours by means of the
\emph{marching squares} algorithm~\cite{Lorensen1987}, whereby the contours are
discretized. The precision level of the discretization is
improved until full convergence of the results. Figure~\ref{fig:waves} illustrates typical
results of numerical diagonalization giving the Fermi-level contours
(inset, see also Fig.~\ref{setup}) and
the wavefunction probability density
$|\psi |^2\equiv|\langle z|\psi\rangle|^2$.

Most bulk states form closed contours located at one of the valleys.  Additionally there is the special contour
that connects the valleys by wrapping around them.
In between  it consists of surface states but
at places where the contour touches the bulk contours, the states are
delocalized --- we call them chiral bulk states, since they are
unidirectional, moving parallel or antiparallel to the intrinsic magnetization
(here  $x$ direction) depending on the valley. While the number of
 bulk contours $\sim k_F W/\pi$ increases with the width,
 there is always only a single contour that contains the
surface and chiral bulk states.

\section{Scattering}

\subsection{Disorder potential}
\label{impscat}

\begin{figure}
\includegraphics[width=\linewidth]{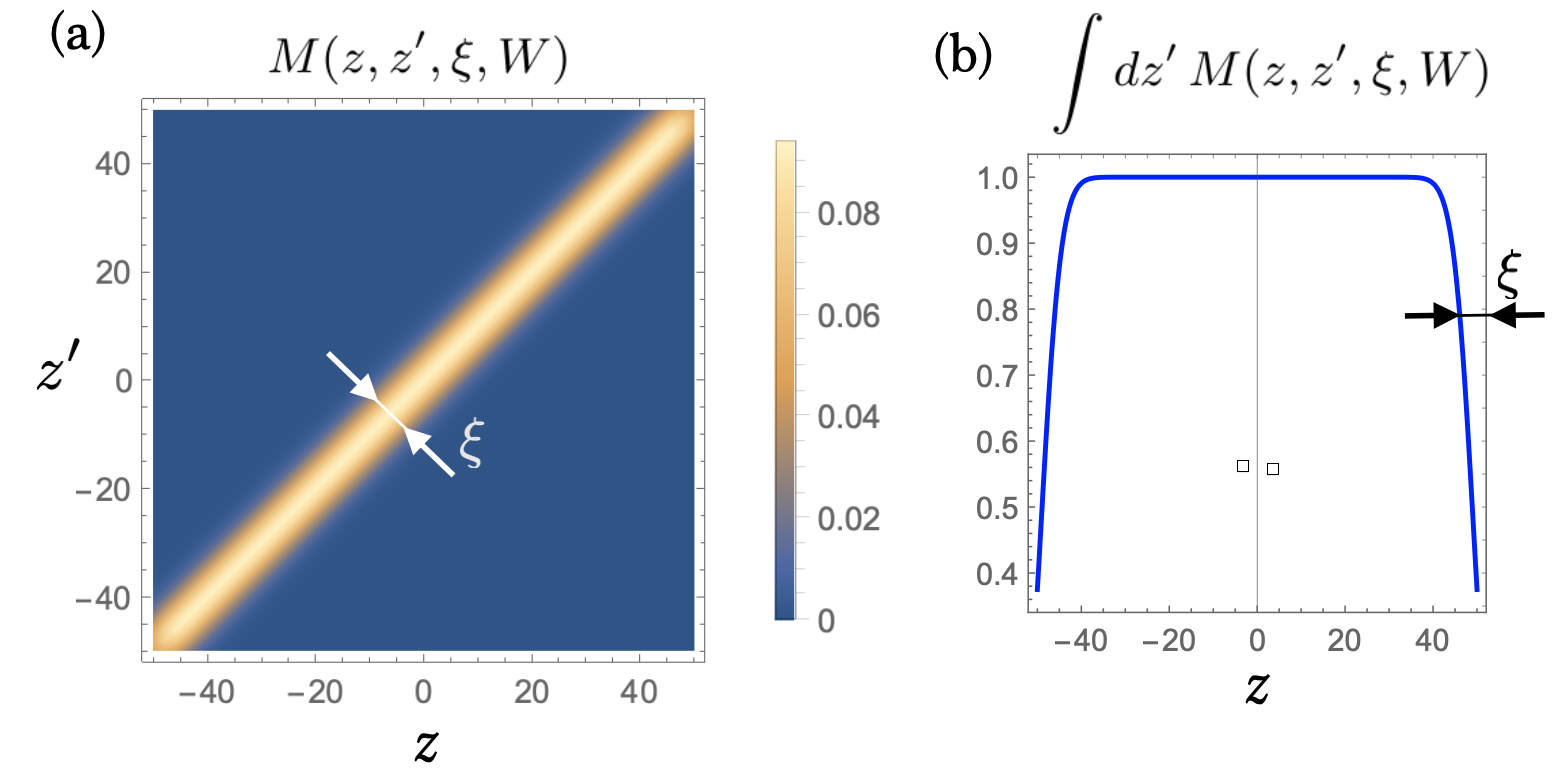}
\caption{Plot of the function $M(z,z',\xi,W)$ (a) and
$\int dz'\, M(z,z',\xi,W)$ (b) at $\xi=3$ and
$W=100$. The width of the diagonal peak in (a) and the region of
reduced weight in (b) are both set by $\xi$. }
\label{fig:mfct}
\end{figure}

We model the disorder  by static Gaussian potentials,
\begin{align}
V &= \sum_\alpha U_\alpha \phi(\vx-\vx_\alpha),&
\phi(\vx) &= e^{-|\vx|^2/2\xi^2},
\end{align}
where the sum runs over the Gaussian's with a
characteristic width $\xi$, random and
uncorrelated potential magnitudes
$U_\alpha \in [-\delta,\delta]$, and random
positions $\vx_\alpha$.

The disorder potential enters the BE~\eqref{BE1} in the form
of the scattering rate between two energy eigenstates
$Q(\kk, \kp) = 2\pi \delta(E_\kk-E_F)q(\kk, \kp)$, where
\begin{equation}
    q(\kk, \kp) = \left\llangle
    \vert\langle\psi_\kk\vert V \vert\psi_\kp\rangle\vert^2
    \right\rrangle
\end{equation}
and $\llangle \dots \rrangle$ denotes disorder average.

In the slab model, as compared to a translation-invariant system, the
scattering rate is not a simple Gaussian as a function of the momentum
difference.
Inserting the impurity potential and averaging over the disorder
configurations within the slab we obtain
\begin{multline}
   q(\kk, \kp)  = \frac{(2\pi\xi^2)^3 \delta^2 n_i}{3 L_x L_y}
    e^{-\xi^2 (\bm{k}-\bm{k}')^2/2}
    \sum_{z,z'}  M(z,z',\xi,W) \\
    \times
    \psi_\kp^\dagger (z) \psi_\kk^\pd (z)
    \psi_\kk^\dagger (z') \psi_\kp^\pd (z')
\label{scatrate}
\end{multline}
where $ \psi_\kk^\pd (z) = \langle z| \psi_\kk\rangle$ and $n_i$ is the impurity
concentration. A detailed derivation can be found in the Appendix. The function $M(z,z',\xi,W) $ is given by
\begin{multline*}
M(z,z',\xi,W) = \frac{ e^{-(z-z')^2/4\xi^2}}{4\sqrt{\pi}\xi} \\
\times
    \left[
    \mathrm{erf}\left(
    \frac{W + z+z'}{2\xi}
    \right)
    +
    \mathrm{erf}\left(
    \frac{W - z-z'}{2\xi}
    \right)
    \right],
\end{multline*}
where the error function is defined as
$\mathrm{erf}\left(x\right)=(2/\sqrt{\pi})\int_0^xe^{-t^2}\mathrm{d}t$. As
illustrated in Fig.~\ref{fig:mfct}, $M(z,z',\xi,W) $ is mainly
the $(z-z')$ dependent part of the Gaussian impurity potential, which
magnitude however reduces by approximately a factor of 2 at the slab surfaces,
where the possible impurity positions obviously fill only half of the space.
Another effect of the finite size is that the $z$ dependence of the
wavefunctions is not plane-wave like, hence the sum in~\eqref{scatrate}
does not reduce to a Fourier transformation. In particular, note that
the wavefunction factor in~\eqref{scatrate} strongly suppresses the
scattering rate between surface states of opposite surfaces
when $\xi$ and the penetration depth of the surface states are both
much smaller than $W$ due to
a vanishing overlap between the surface states.

In the limit $\xi\ll 1$ one obtains
$M(z,z',\xi,W) = \delta(z-z')$, in which case
\begin{multline*}
q(\kk,\kp)
\xlongrightarrow{\xi\ll 1}
\frac{(2\pi\xi^2)^3 \delta^2 n_i}{3 L_x L_y}
e^{-\xi^2 (\bm{k}-\bm{k}')^2}
\\
\times
\sum_{z} \,
|\psi_\kp^\pd (z)|^2 |\psi_\kk^\pd (z)|^2.
\end{multline*}

Despite the rather complex form, the impurity scattering is fully
determined by two parameters
--- the real-space impurity width $\xi$ and
the overall impurity strength
(set by $\delta^2n_i$), the latter will be in the
following quantified by the average mean free path $l$, defined below.
The impurity width $\xi$ essentially sets the
momentum-space range of most scattering processes
to $|\bm{k}-\bm{k}'|\lesssim \xi^{-1}$, hence for a large
value of $\xi$ scattering between states that are far apart
from each other in the in-plane momentum $\bm{k}$ is exponentially suppressed $\sim \exp[-\xi^2(\bm{k}-\bm{k}')^2]$.

\subsection{Scattering lengths}
\label{sls}

In the presence of surface and bulk states it is interesting
to quantify averaged scattering rates and scattering probabilities between
different types of states, which will be
helpful to understand the numerical transport results.

To quantify the overall strength of impurity scattering, we define
the averaged mean free path
\begin{equation}
 l=\left\langle \, |\vk| \left(\sum_{\kp} \delta(E_\kk-E_F) q(\kk, \kp)
  \right)^{-1}\,\right\rangle,
\end{equation}
where the Fermi-surface average is given by
\begin{align}
\langle\dots\rangle &=
\frac{1}{N} \sum_\kk \delta(E_\kk-E_F)(\dots );
&
N &= \sum_\kk \delta(E_\kk-E_F),
\label{eq:fs_average}
\end{align}
and $N$ is the number of states at the Fermi level. Note that for
a nearly constant  $|\vk|$ that we have, the mean free path is
inversely proportional to the total scattering probability.

To quantify the scattering probability between different types of
states $i\in[b,\;  s]$, where $b$ ($s$) denotes bulk (surface) states, we define the
scattering length
\begin{equation}
 l_{ij} =\left\langle| \vk| \left(\sum^j_{\kp} \delta(E_\kk-E_F) q(\kk, \kp)
  \right)^{-1}\right\rangle_i
  \label{lij}
\end{equation}
where the sum runs over the type $j$ of states and the averaging is analogous to
\eqref{eq:fs_average} but only over type $i$ of states ($\sum_\kk \to \sum^i_\kk$).

We now want to determine the dependence of the
scattering probability on the width $W$ of the slab
in two limiting cases (i) the number of
 bulk states being much larger than that of surface states,
$N\gg N_s$ and (ii) the number of states
being dominated by surface states,
$N \sim N_s$. In regime (ii)
the number of bulk states $N_b=N - N_s$, is given
by the number of chiral bulk states. Thus the
scaling of $N_b$ with the slab width reads
\begin{equation}
N_b \sim\begin{cases} W & \text{(i)}\\ 1 & \text{(ii)}.\end{cases}
\label{dosw}
\end{equation}

The scaling of the scattering rate~\eqref{scatrate} is governed by the
 $z$ dependence of the wavefunctions.
A normalized surface state with a penetration depth $\lambda \sim \beta^{-1} \sim 1$ and
a normalized bulk state are of the form
\begin{align}
\psi_s(z) &= \sqrt{\frac{2}{\lambda}}\, e^{-z/\lambda},
&
\psi_b(z) &= 1/\sqrt{W},
\end{align}
respectively. Consequently, for
$W \gg \lambda  $ (which we always consider),
we can estimate the scaling of the scattering
rate between the different types of states
as
\begin{equation}
q(\kk,\kk') \sim
    \begin{cases}
        1 & \text{same surface} \\
        \frac{1}{W} &  \text{bulk-surface, bulk-bulk} \\
        0 & \text{opposite surfaces}.
    \end{cases}
\end{equation}
From this,~\eqref{dosw}, and~\eqref{lij}    the width
dependence of the scattering probabilities summarizes to
\begin{equation}
\begin{tabular*}{0.9\columnwidth}{@{\extracolsep{\fill}}lllll}
\hline
       & $l_{bb}/l$  & $l_{ss}/l$  & $l_{sb}/l$  & $l_{bs}/l$  \\ \hline
(i)   & $1$ 	    & $\tfrac{N_b}{N_sW}\sim 1$        &  $1$        &  $\tfrac{N_b}{N_s}\sim W$      \\
(ii)   & $W\tfrac{N_s}{N_b}$ 	    & $1$        &  $W\tfrac{N_s}{N_b}$       &  $W$ \\
\end{tabular*}
 \label{lscaling}
\end{equation}
Most importantly,  in the regime (i)
the scattering probability from bulk to surface ($\propto l^{-1}_{bs}$)
is a factor $W$ smaller than other scattering probabilities,
due to the ratio of the number of bulk states to surface states, which is
large and increases with $W$.
In the regime (ii) instead,
the number of bulk states (consisting only of the chiral bulk states)
does not depend on $W$. The peculiarity of this regime is that surface states
scatter most probably within the surface states, which is due to the larger overlap
of surface wavefunctions.

\section{Numerical results}
\label{sec:cond}

We calculate the nonequilibrium occupation function 
\eqref{ansa} at zero temperature by numerically solving 
Eq.\ \eqref{BE1} with respect to the transport length 
on the basis of the numerical solution of the
discretized slab 
spectrum discussed in Section III  \cite{zenodo}. The 
nonequilibrium occupation function 
(or, equivalently, the transport length) determines the 
conductivity, given 
in Eq.\ \eqref{sigma}, and the
valley polarization, 
defined in \eqref{chi} below.

\subsection{Conductivity}

We consider the conductivity  in units of the standard Drude estimate
given by the mean free path $l$,
the density  of states at the Fermi level $n=N/V$, and
the Fermi velocity $v\approx t /\hbar$,
\begin{equation}
\sigma_0 = \frac{e^2 n\, l \,v }{3}.
\label{s0}
\end{equation}
This is the result one would expect to find for a system with point-like
impurities ($\xi \to 0$) and only bulk states arranged in
form of a spherical Fermi surface.

\begin{figure*}
\includegraphics[width=0.95\linewidth]{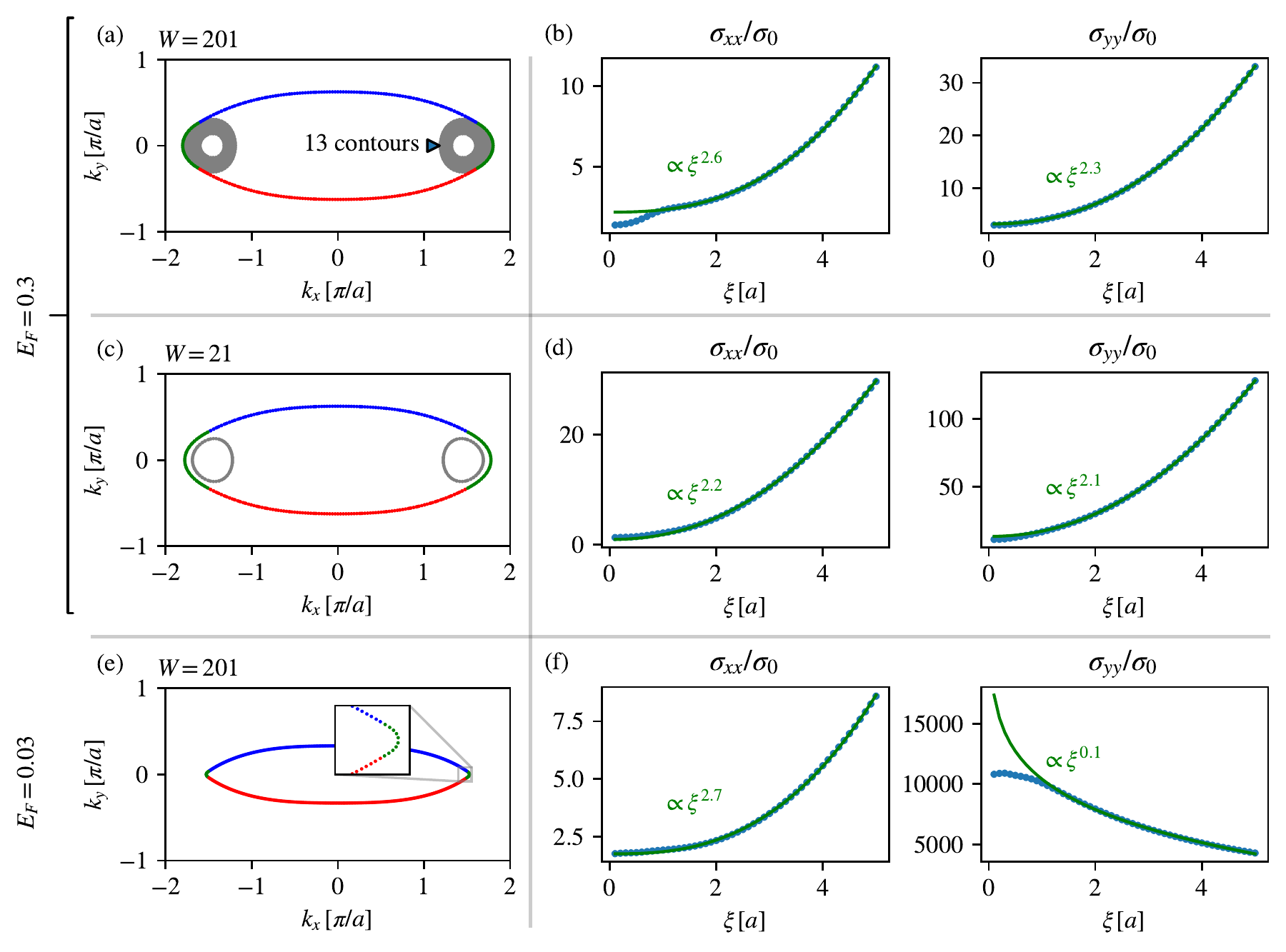}
\caption{
Impurity-range dependence of the conductivity. (a),(c),(e): Discretized Fermi-level contours
for the three parameter choices ($E_F=0.3\, t,\; W=201$), ($E_F=0.3\, t,\; W=21$),
and ($E_F=0.03\, t,\; W=201$). 
The number of finite-size induced states 
(surface states and chiral bulk states), $N_s+N_c$,
relative  to the total number of states $N$
is (a) $(N_s+N_c)/N = 0.14$, (c) $(N_s+N_c)/N = 0.68$, 
and (e) $(N_s+N_c)/N = 1$.
Color indicates
surface states at the upper (blue) and lower (red) surfaces, bulk states (gray), and the chiral bulk states (green),
cf.\ Fig.~\ref{fig:waves}. (b), (d), (f): Conductivity
as a function of the impurity
width $\xi$. The green curves show
 power-law fits, the exponent is indicated in the plot. Other parameters are
$\beta =1.5,\; \; \mu_b=0.3\, t$.}
\label{fig:xidep}
\end{figure*}

The dependence of the slab conductivity on
the width of the impurity
potential $\xi$ is summarized
in Fig.~\ref{fig:xidep}.
For $\xi >1$ the conductivity is well fitted by a power-law dependence on $\xi$,
with an exponent between 2 and 3. An
exception is found for $\sigma_{yy}$ at $E_F\ll t$, which shows a weak
$\xi$ dependence at a
strongly enhanced conductivity at all $\xi$.
In total, the magnitude of the conductivity,
especially in the direction of motion of Fermi arcs, may be enhanced by several
orders of magnitude, either due to a wide impurity range, or if the Fermi energy
is close to the Weyl points.

\begin{figure*}
\includegraphics[width=0.95\linewidth]{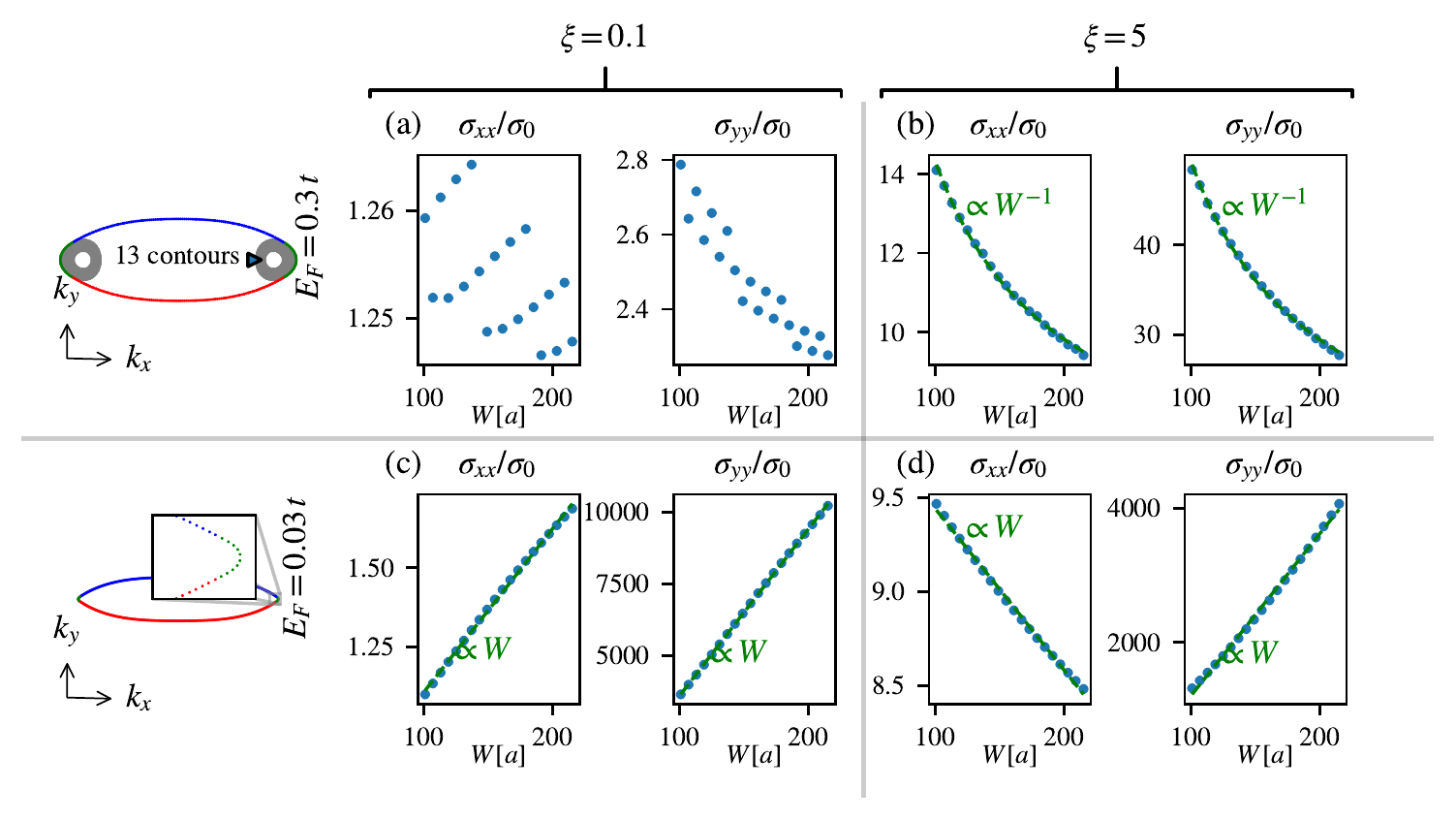}
\caption{ Width dependence of the conductivity at 
 (a) $E_F= 0.3\, t$, $\xi =0.1$, (b)  
$E_F= 0.3\, t$, $\xi =5$, (c) 
 $E_F= 0.03\, t$, $\xi =0.1$, and
$E_F= 0.03\, t$, $\xi=5$. Other parameters are
$\beta =1.5,\;\mu_b=0.3\, t$. The
dashed lines in (b) indicate $W^{-1}$ dependence and in (c) and (d) linear dependence.}
\label{fig:width}
\end{figure*}

To gain further insight, in Fig.~\ref{fig:width} we consider the width dependence of the conductivity.  Figure~\ref{fig:width}(a)
shows that in the case of a large number of bulk states and point-like impurities,
the conductivity is nearly independent of $W$ and is close to $\sigma_0$ ---
in this regime the slab thus resembles a conventional
metal.
At a large $\xi$, however, the conductivity
enhancement decreases antiproportional to the width, indicating
that the conductivity enhancement at large $E_F$
is related to the presence of the confinement-induced
surface and
chiral bulk states, which number, relative to the total number of states $N$,
is antiproportional to $W$. At small $E_F$, however, when there is only one
Fermi-level contour mainly consisting of surface states, the total number of states
is nearly independent of
$W$. In this case, the large conductivity in the direction of motion
of surface states, $\sigma_{yy}$, linearly \emph{increases} with the width $W$.

\begin{figure}
\includegraphics[width=0.95\linewidth]{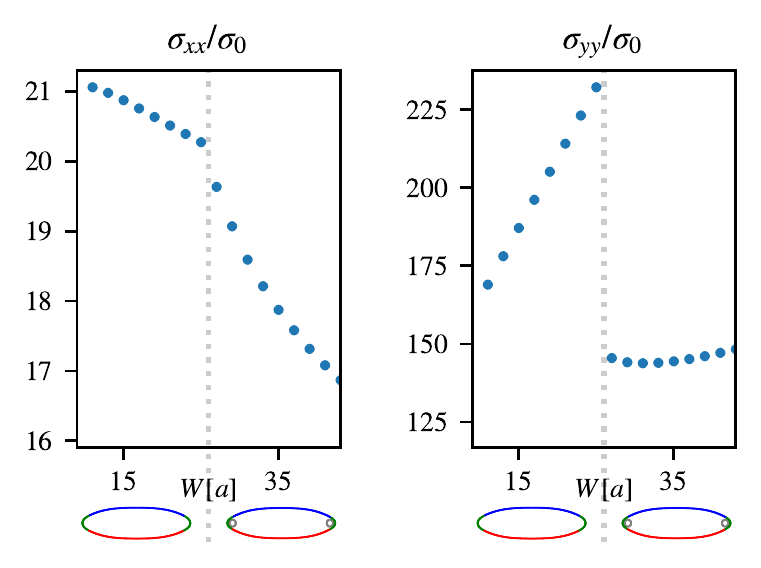}
\caption{ Width dependence of the conductivity at $\xi =5$ and
$E_F=0.15\, t$, $\mu_b = 0.3\; t$. The transition from a single contour to three
contours is indicated by the dotted lines and the contour-plots below the $W$ axis. }
\label{fig:width2}
\end{figure}

To explore more carefully the transition from an increasing to a decreasing $W$
dependence, in Fig.~\ref{fig:width2} we plot
the conductivity at smaller width, when the first bulk
contours appear. This plot shows that the strong $W$ enhancement of the
surface conductivity requires
the normal bulk contours to vanish, which happens if
$W k_F/\pi \lesssim 1$.
As soon as at least one normal bulk contour appears, the conductivity jumps to a
significantly lower value and becomes
decreasing in $W$.

\subsection{Valley polarization}

Besides the conductivity it is interesting to explore the non-equilibrium occupation
difference of the valleys, which occurs when the electric field
points along the valley separation, $\vE=\hat{\bm{x}} E$.
The average occupation of the valley at $k_x=\pm \beta$  is given by
$\sum^\pm_\kk g_\kk/( N_b/2) $, where the sum runs over all bulk states
at the valley $\pm$, and $N_b/2 = \sum^\pm_\kk$ is the number of those states.
We quantify the valley polarization by the difference
of the valley occupations
relative to the standard occupation difference of states
$\delta(E_F-E_\kk) eEl$ due to the mean free motion in the electric field,
\begin{equation}
\begin{aligned}
\chi & \equiv
 \dfrac{\sum^+_{\kk}g_\kk}{\sum^+_{\kk}\delta(E_F-E_\kk) eE l}
-\dfrac{\sum^-_{\kk}g_\kk}{\sum^-_{\kk}\delta(E_F-E_\kk)eE l} \\
&= \frac{\langle\Lambda^x_\kk\rangle_+-\langle \Lambda^x_\kk\rangle_-}{l},
\end{aligned}
\label{chi}
\end{equation}
where in the second line we used Eq.~\eqref{ansa} and defined the average over
valley bulk states $\langle\dots\rangle_\pm=
\sum^\pm_{\kk}\delta(E_F-E_\kk)\dots /\sum^\pm_{\kk}\delta(E_F-E_\kk)$.

\begin{figure}
\includegraphics[width=0.95\linewidth]{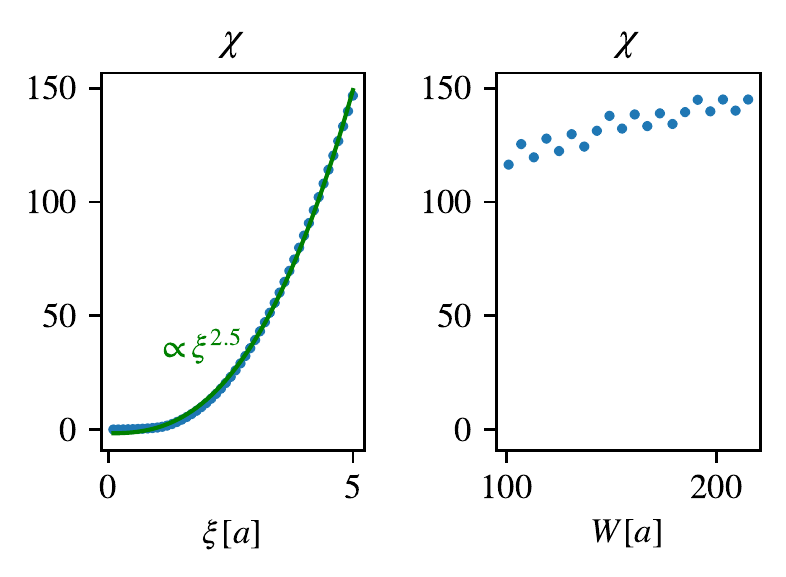}
\caption{ Valley polarization $\chi$, defined in Eq.~\eqref{chi}. (a) Dependence
of $\chi$ on the impurity range  $\xi$ at a width $W=201$. (b)
Width dependence of $\chi$ at $\xi=5$. Other parameters are $\mu_b=0.3\, t$, $\beta=1.5$, and
$E_F=0.3\, t$. }
\label{fig:valleyp}
\end{figure}

A representative result for the impurity range and width dependence of
 the valley polarization is shown in Fig.~\ref{fig:valleyp}.  The valley
polarization shows the power-law $\xi$ dependence, similar to the
conductivity. However, unlike for the conductivity, there is no significant width
dependence at large $\xi$ (here already at $\xi\gtrsim 1$)
in the numerically
accessible width range. This is very surprising as it seems to imply a presence of valley polarization for sufficiently
large impurity widths in arbitrary large systems, contradicting previous predictions based on infinite-system calculation \cite{Armitage2017}.
 Below we will show that the valley polarization 
in fact does decay but for width above
$\gtrsim l\ e^{(\xi\Delta k)^2}$, which becomes 
exponentially large for $ \xi\Delta k \gtrsim 1$.

\section{Discussion}

The numerical results of the previous section show enhancements of the slab conductivity
by several orders of magnitude (compared to the
expectation for a conventional metal~\eqref{s0})
and a substantial valley polarization in a wide region of the parameter space.
The characteristic
dependencies on $\xi$ and $W$ allow
to identify the main mechanisms of these effects, which we now systematically discuss.

\subsection{Impurity-range dependence}
\label{toymodel}

When the impurity range $\xi$ increases, the scattering rate
\eqref{scatrate} between two
countermovers separated by $\Delta k$ becomes exponentially suppressed,
$\sim \exp[-(\xi\Delta k)^2]$. The transport length
does not inherit the exponential enhancement though, since relaxation happens
via multiple small-angle scattering processes. To illustrate this, we consider a toy model
 of a closed chain of ${\cal N}$ states labeled
$i\in [1,{\cal N}]$ with arbitrary
velocities $\bm{v}_i$.
The BE~\eqref{BE1} is of the form
 (summation over repeated indices assumed)
\begin{equation}
\bm{v}_i = M_{ij} \LL_j,
\label{beqtoy}
\end{equation}
where $M_{ij} = \delta_{ij} \sum_k q_{ik} -q_{ij} $
is given by the scattering rates between states, $q_{ij}$.
We assume scattering only between the
nearest neighbors, with the rate
$q_\mathrm{nn}$, and direct scattering between countermovers with the
rate $q_\mathrm{d}$, in which case the matrix $M$ becomes
\begin{equation}
M_{ij} = (2q_\mathrm{nn} + 2 q_\mathrm{d}) \delta_{ij}
- q_\mathrm{nn}(\delta_{ij+1} + \delta_{i,j-1}),
\end{equation}
where we used that the direct-scattering part of $q_{ij}$ cancels when multiplied with $\LL$.
For the nearest-neighbor part $M_\mathrm{nn}$ of $M$ there is a left pseudo-inverse,
\begin{equation}
 P_{ij} = \frac{1}{q_\mathrm{nn}}\frac{|i-j|\left( |i-j| - {\cal N} \right)}
 {{2 \cal N}},
\end{equation}
so that $PM_\mathrm{nn} \LL = \LL$ (note $\sum_i \LL_i = 0$ due to particle
conservation).
With its help, the full solution of~\eqref{beqtoy} becomes
 \begin{equation}
 \LL =   \left(1+
 2 q_\mathrm{d} P\right)^{-1} P\, \bm{v}.
  \label{toyl1}
 \end{equation}
We first ignore direct scattering. From the form of $P$, it is clear that
the solution can depend on ${\cal N}$ to maximally the power ${\cal N}^2$.
In particular, for a circular velocity arrangement
$\bm{v} = \{ - \cos [2\pi (i-1)/{\cal N }] ,\; \sin [2\pi (i-1)/{\cal N}] \} $
the vector mean free path for ${\cal N} \gg 1$ assumes the value
\begin{equation}
\LL = \frac{1}{q_\mathrm{nn}}\frac{(\tilde{\xi} \Delta k)^2}{4}  \bm{v},
\end{equation}
where we have written the number of nearest neighbors
as  ${\cal N} = \pi\, \tilde{\xi} \Delta k$, in terms
of  the spacing between
nearest neighbors $\tilde{\xi}^{-1}$ and the distance between
countermovers $\Delta k$.

Considering the full result with direct scattering
in Eq.~\eqref{toyl1}, we
see that the nearest-neighbor scattering contributions to $\LL$ dominate
 as long as
\begin{equation}
\frac{(\tilde{\xi} \Delta k)^{p} }{q_\mathrm{nn}}\ll \frac{1}{q_\mathrm{d}},
\end{equation}
where the exponent $p$ depends on the velocity arrangement.
In the opposite limit we instead obtain
\begin{equation}
\LL = \frac{1}{2 q_\mathrm{d}}   \bm{v}.
\end{equation}
Transferring this insight to the Weyl slab model, the distance between
the nearest neighbors $\tilde{\xi}^{-1}$ corresponds to
 $ \xi^{-1}$. The mean free path corresponds to the inverse diagonal
 of $M$ times velocity, $l\sim v/(q_\mathrm{d}+q_\mathrm{nn})$. Interpolating
 between the two regimes of dominant direct scattering and nearest-neighbor scattering
 the  typical values of the transport length may be well estimated as
 \begin{equation}
 \bar{\Lambda} = l\,\left[1+(\xi \Delta k)^{p}\right].
 \label{aLtoy}
 \end{equation}
This explains the general power-law conductivity enhancement with $\xi$ in
Fig.~\ref{fig:xidep}, except for $\sigma_{yy}$ in Fig.~\ref{fig:xidep}(f), which we discuss separately.

\subsection{Valley polarization}

The averaged occupation of the two valleys $\pm$  can be expressed as
\begin{equation}
    \mu_\pm = \frac{\sum^\pm_{\kk} g_\kk}{N_b/2},
\end{equation}
where $\sum^\pm_{\kk}\delta(E_F-E_\kk) =N_b/2$.
The occupation difference is related to the valley polarization $\chi$ defined in Eq.~\eqref{chi},
\begin{equation}
    \mu_+-\mu_-  = \chi\, eE_xl.
\end{equation}
A single valley has a total velocity  in the $x$ direction  ---
the unbalanced velocity of the chiral states
$vN_c/2$, where $N_c$
is the total number of chiral states ($N_c/2$ in each valley).
An electric field in the $x$ direction thus pumps charge between the valleys
with the rate $veE_xN_c$,
which in a steady state
must be counterbalanced by scattering.
We can write down a simple balance equation as a condition for
a steady valley occupation,
\begin{equation}
    e N_b \frac{d\mu_\pm}{dt} = 0=  \pm veE_x  N_c
    + eN_b
    \left(\frac{\partial \mu_\pm}{\partial t}\right)_\text{scat},
    \label{cont}
\end{equation}
where we used $N_b\gg N_c$. The time-change of $\mu_+-\mu_-$ due to scattering
is proportional to the occupation difference itself and the scattering probability,
which we quantify by the  scattering length $l_c$, hence
\begin{equation}
    \left(\frac{\partial ( \mu_+-\mu_-)}{\partial t}\right)_\text{scat} =
    -2 \frac{\mu_+-\mu_-}{l_{c}/v}
    =-2 \frac{\chi eEl}{l_{c} /v}.
\end{equation}
Together with~\eqref{cont}  we obtain
\begin{equation}
\chi = \frac{N_c}{N_b}\,\frac{l_{c} }{l}.
\end{equation}
For a point-like disorder potential ($\xi \to 0$) the ratio $l_c/l$ goes to one
and the valley polarization is small. For larger $\xi$, direct scattering between
the valleys becomes strongly suppressed and the relaxation of  $ \mu_+-\mu_-$
must go via surface states. Thereby the relaxation along the arcs and the scattering
from surface to bulk is much faster than the scattering from bulk to surface,
see~\eqref{lscaling}. The scattering length $l_c$ is thus set by
the bulk-surface scattering
length $l_{bs}$, which is proportional to the ratio $N_b/N_s$ so that
\begin{equation}
    \chi \propto \frac{N_c}{N_s}.
\end{equation}
This explains the surprising result that the valley polarization does not depend on the
width, as seen numerically in Fig.~\ref{fig:width2}. This is surprising since
the origin of the valley polarization are
the chiral bulk states, which number $N_c$ is a factor $\propto W$ smaller
than the total number of bulk states $N_b$. The explanation is that the valley relaxation
 also becomes suppressed $\propto W^{-1}$ since the probability to
 scatter into Fermi arcs
decreases with an increasing number of bulk states.
For a much larger width, when the probability of relaxation
via Fermi arcs becomes smaller than the probability of relaxation via direct inter-valley scattering,  the valley polarization will ultimately go to zero like $W^{-1}$.
In case of a Gaussian potential, the amplitude of direct inter-valley cattering is however
exponentially suppressed
$\sim \exp[-(\xi\Delta k)^2]$ so that the width independence can easily extend to arbitrary macroscopic sizes
for realistic values of the impurity width
and the cone separation $\xi \Delta k \gtrsim 1$.

Regarding the  strong enhancement of valley
polarization with $\xi$, one is tempted to understand
it as a consequence of an increasing number of
scattering events needed for a relaxation
along the Fermi arcs. However, the contribution of such a process
to $\chi$ would enter in the form $ l_{ss} (\xi\Delta)^p /l \propto 1/W $, with
a clear $1/W$ dependence, which we do not observe.
It rather must be the suppression of the bulk-surface
scattering probability with $\xi$, which is
plausible in view of the typical effect of an
increasing $\xi$ to increase the relaxation time.
We note, however, that
the simplified analytical calculation of Section~\ref{toymodel} does not
apply since in this case the valleys correspond to
two-dimensional pools of states,  while the model of Section~\ref{toymodel}
only  considers  one-dimensional chains.

\subsection{Conductivity in case of a large number of bulk states}

For bulk states (excluding chiral bulk states) the small separation of countermovers in momentum space
$\sim k_F$ makes their
transport length close to the mean free path $l$ and not significantly enhanced with $\xi$.
We thus approximate the current contribution of  bulk states as
\begin{equation}
    \bm{j}_{n} = \sigma_0\bm{E}.\label{jnn}
\end{equation}
For an electric field in the $y$ direction there is additionally the contribution
of surface states, which transport length is mainly set by
surface-surface scattering, enhanced by $\xi$ according to~\eqref{aLtoy},
\begin{equation}
    \bm{j}_{s} =  e^2 n_\mathrm{s}v\, l_{ss} (\xi\Delta k)^p \bm{y} E_y,
\end{equation}
where $n_s=N_s/V$ is the density of surface states.

The current contribution of chiral bulk states is negligible compared to \eqref{jnn}, except if the valley polarization becomes large, in which case
\begin{equation}
    \bm{j}_c=  e v \hat{\bm{x}} n_c(\mu_+- \mu_-)/2
    = e^2 v \hat{\bm{x}} n_c\chi l \, E/2,
\end{equation}
where $n_c=N_c/V$ is the density of chiral bulk states.
Adding all the current contributions,
we obtain
\begin{align}
    \frac{\sigma^{(i)}_{xx}}{\sigma_0} &= 1+\frac{3}{2}\frac{N_c}{N_b} \chi,   \\
    \frac{\sigma^{(i)}_{yy}}{\sigma_0} &= 1+3\frac{N_s}{N_b} \frac{l_{ss} (\xi\Delta k)^p }{l}
    \sim 1+ \frac{1}{W}  (\xi\Delta k)^p ,
\end{align}
where we again used~\eqref{lscaling}.
This rough estimate is in qualitative agreement with the numerical results.
Note that $N_b$ increases proportional to $W$, which explains the $W$ dependence of
$\sigma_{xx}$ and $\sigma_{yy}$ in  Fig.~\ref{fig:width}(b).
We now understand that the enhancement of $\sigma_{xx}$ and $\sigma_{yy}$
are mainly due to chiral bulk states and surface states, respectively.

In conventional metals the
conductivity is width independent
as long as the width is much larger than the mean free path;
when the width
becomes smaller, the conductivity tends to
decrease due to additional scattering at boundaries. Our work shows that in Weyl
semimetals away from  charge neutrality the opposite
trend of increasing conductivity with shrinking the width may occur due to chiral bulk states and surface states. A
qualitatively similar width dependence
may occur also in the regime $l\ll W$ (we consider the opposite limit
$l\gg W$), which has been considered in
Ref.~\onlinecite{Breitkreiz2019}.
Observations of enhanced  conductivity for reduced widths
have been reported in Ref.~\onlinecite{Zhang2019},
where the  Weyl semimetal nanobelts,
according to estimates of the mean free path, are presumably in the
regime of our work, $l\gg W$, or in the crossover regime  $l\sim W$.

\subsection{Conductivity, small number of bulk states}

We now come to the case (ii) when the number of states
is dominated by surface states, while in the bulk only
chiral bulk states are present.
The  conductivity in the $y$ direction can be written in the form
\begin{equation}
    \frac{\sigma^{(ii)}_{yy}}{\sigma_0} =
    1+\frac{N_s}{N}\frac{\Lambda_s}{l},
\end{equation}
where $N_s\approx N$ is the number  and $\Lambda_s$ the
transport length of surface states.
Scattering within surface states at the same surface does not
lead to relaxation of motion in the $y$ direction
since the average velocity $v_y$ of those states is not zero.
Countermoving surface states, on the other hand, have no overlap with each other,
direct scattering between them is blocked.
The relaxation of Fermi arc states must thus go via
the small number of chiral bulk states, so that  $\Lambda_s$ is  set by the surface-bulk
scattering probability,  $\Lambda_s \sim l_{sb}$. In Section~\ref{sls}
we found $l_{sb}/l \sim W N_s/N_c$---Eq.~\eqref{lscaling}---which leads to
\begin{equation}
    \frac{\sigma^{(ii)}_{yy}}{\sigma_0} \sim W \frac{N_s}{N_c}.
\end{equation}
Both
factors are of order $100$ for parameters in Fig.~\ref{fig:xidep}(f),
which explains the large magnitude. Also the $W$ dependence in Fig.~\ref{fig:width}
is consistent since both $N_s$ and $N_c$ are $W$ independent in this case.

For large $ \xi$, surface-bulk scattering becomes limited to small regions at the nodes and
the full relaxation must involve nearest-neighbor scattering along the Fermi arc. According
to Section~\ref{toymodel}, the latter
should elongate the full transport length by an additional $l_{ss}(\xi\Delta k)^p $.
For the considered parameters, $l_{sb}$ is much larger than this additional part since
 $(\xi\Delta k)^p \ll  WN_s/N_c$, which explains the weak $\xi$ dependence in  Fig.~\ref{fig:xidep}(f).

The transition from (ii) to (i) spoils
the strong enhancement of $\sigma_{yy}$ in two ways: First, the ration $N_s/N_c$ changes to  $N_s/(N_c+N_b)$
and thus becomes smaller and
second, according to Eq.~\eqref{lscaling}, $l_{sb}/l \sim W N_s/N_b \to 1$
is no longer width dependent,
which in Fig.~\ref{fig:width2} explains the jump and the change of slope.

The conductivity in the $x$ direction is also governed by
the dominant number of surface states.
Relaxation however happens via scattering within the same surface,
since $v_x$ averages to zero at each surface separately.
Since $l_{ss}/l \sim 1$, the conductivity
in the $x$ direction is not significantly enhanced.

\section{Conclusion}

In conclusion, we have studied linear-response properties of a finite
Weyl semimetal slab (width $W$)
in the presence of long-ranged disorder
(disorder potential width $\xi$).
Our work highlights the remarkable property of Weyl semimetals to
realize valleys of opposite chirality that are well separated in momentum space
($\Delta k$) and continuously connected only via surface states. For a Fermi energy that is not exactly at the Weyl nodes, the surface states occur together
with confinement-induced
chiral bulk states. In the presence of an electric
field parallel to cone separation they allow to violate chiral charge conservation even without an external magnetic field.
This peculiarity stabilizes an anomalous valley polarization
at zero magnetic field.
If the potential width is
substantially larger than the
inverse separation of valleys,
the valley polarization persists up to
very large slab width.
This is explained by the fact that direct inter-valley scattering is strongly
suppressed and the
relaxation must go via Fermi arcs, which
is however also increasingly ineffective owing to their vanishing density with
an increasing width. The resulting
width independence of the
confintenment-induced valley polarization persists
up to a width, for
which relaxation via direct inter-valley
scattering becomes more effective
than relaxation via Fermi arcs.
For Gaussian-type disorder potentials
this maximum width is exponentially enhanced by
$\exp[(\xi\Delta k)^2]$ and can thus easily reach
macroscopic length scales at realistic values of
cone separation $\Delta k$ and
inpurity-potential widths $\xi$.

The valley polarization and Fermi-arc surface states lead
to a conductivity enhancement which increases 
with an increasing width of the disorder potential and 
and a decreasing width of the slab 
($\sigma \propto 1/W$).
Moreover, if the Fermi energy is reduced towards charge neutrality
such that normal bulk states vanish completely,
the conductivity in the direction of motion of surface states becomes
strongly enhanced because relaxation of surface states can only go via
bulk states which number becomes strongly reduced.

Methodologically our work performs first steps in the application of the weak-disorder transport formalism to a multilayer system
with a large ($\gtrsim 100$) number of layers and consequently a
similarly large number of bands in the in-plane Brillouin zone.
The numerical code 
\cite{zenodo} is designed to be easily 
applicable to an arbitrary lattice model and
 can thus be used to explore 
in detail the
confinement-induced valley polarization in various
Weyl-metal models. In this work,  
we introduced the formalism by considering 
a minimal two-Weyl-cone
model. We find that the qualitative aspects of the
valley polarization  are robust to lattice details such as 
boundary potentials (which give the Fermi arcs
a finite curvature) or velocity anisotropy of the Weyl cones.
Our analytical discussion shows that the 
valley polarization depends on the mere presence 
of chiral bulk states and surface states (which is 
topological) and the ratio of the inverse separation of Weyl
cones vs.\ the width of the scattering potential, which 
 explains the robustness of this effect.  
 
An interesting application of the introduced 
tools is to consider lattice models of existing 
Weyl semimetals. For the case of several pairs of Weyl nodes that are sufficiently separated in momentum space, 
such as in the TaAs material family, we expect 
 valley polarization  to 
occur 
 in each pair which cone sepration 
aligns with the electric field, similarly to the two-cone case. The reason is that due to 
the large pair separation, scattering between pairs
should be negligible compared to the intervalley scattering 
within a single pair, making each 
pair independent and thus reduce the problem to  the 
two-cone case. 

General limitations of the 
introduced numerical tools 
are the restriction to slab width being smaller than the 
mean free path and the restriction to 
the leading order in the
disorder potential. Both the fate of confinement-induced 
effects for larger widths as well as corrections of higher order in the disorder potential, which are known to 
start with Berry phase effects \cite{Xiao2010}, constitute interesting directions to extend this formalism.

\subsection*{Data availability}
All the code and data used to produce the reported results is available in Ref.~\cite{zenodo}.

\subsection*{Author contributions}
M.B. formulated the project idea, developed the theory with input from N.B. and A.A., performed and analyzed numerical calculations, and developed the analytical model. P.M.P.P. developed the numerical code with input from N.B., A.A., and M.B., and performed and analyzed numerical calculations. The manuscript was written by M.B. with input from P.M.P.P. and A.A.

\begin{acknowledgments}
    This research was supported by the European Union Horizon 2020 research and innovation programme under Grant Agreement No.~824140, Grant No.~18688556 of the Deutsche Forschungsgemeinschaft (DFG, German Research Foundation), ERC Starging Grant 638760, and the Netherlands Organisation for Scientific Research (NWO/OCW), as part of the Frontiers of Nanoscience program.
\end{acknowledgments}    

\bibliography{library}

\clearpage
\setcounter{figure}{0}
\setcounter{page}{0}
\setcounter{section}{0}
\setcounter{equation}{0}
\onecolumngrid
\renewcommand{\thepage}{SM\arabic{page}}
\renewcommand{\theequation}{S\arabic{equation}}
\renewcommand{\thefigure}{S\arabic{figure}}
\renewcommand{\thesection}{S\arabic{section}}
\renewcommand{\bibnumfmt}[1]{[S#1]}
\renewcommand{\citenumfont}[1]{S#1}

\begin{center}
\textbf{\large Supplementary Material}
\end{center}

\section*{Derivation of scattering amplitudes}

We consider Gaussian-type static impurity potentials,
\begin{align}
    V &= \sum_\alpha U_\alpha \phi(\vx-\vx_\alpha),&
    \phi(\vx) &= e^{-|\vx|^2/2\xi^2},
\end{align}
where the sum runs over impurities with a
characteristic width $\xi$, random and
uncorrelated potential magnitudes
$U_\alpha \in [-\delta,\delta]$, and random
positions $\vx_\alpha$.

For our transport consideration we consider the
scattering rate
$Q(\kk, \kp) = 2\pi \delta(E_\kk-E_F)q(\kk, \kp)$
between energy eigenstates
$|\psi_\kk\rangle$ and $|\psi_\kp\rangle$, which we calculate using
Fermi's Golden Rule,
\begin{equation}
    q(\kk, \kp) = \left\llangle
    \vert\langle\psi_\kk\vert V \vert\psi_\kp\rangle\vert^2
    \right\rrangle,
    \label{eq:qkkp}
\end{equation}
where the disorder average is defined as
\begin{align}
    \llangle  \dots \rrangle &= \prod_\alpha
    \int_{-\delta}^{\delta}\frac{dU_\alpha}{2\delta}
 \int \frac{d\vx_\alpha}{V}
    (\dots).
    \label{eq:average}
\end{align}
We write the normalized wavefunctions as
\begin{align}
    \langle \vx \rvert \psi_\kk \rangle &=\; \frac{1}{\sqrt{L_xL_y}}e^{i \bm{k}\cdot\rp}\, \psi_\kk(z),
\end{align}
where $\rp = (x,y)$ is the in-plane position and $L_xL_y$ the
in-plane volume and $\psi_\kk(z)$ is the normalized
eigenvector of numerical diagonalization of the lattice model,
 $z$ denoting the discrete sites in the $z$ direction.

The expectation value of the impurity then calculates to
\begin{align}
    \langle \psi_\kp \lvert V \rvert \psi_\kk \rangle &=
    \frac{2\pi \xi^2}{L_xL_y}e^{-\xi^2 (\bm{k}-\bm{k}')^2/2}
    \sum_\alpha  U_\alpha\, \sum_z   \phi(z-z_\alpha) \,
    \psi_\kp^\dagger (z) \psi_\kk^\pd (z).
\end{align}
Inserting into~\eqref{eq:qkkp} and using~\eqref{eq:average},
we obtain
\begin{equation}
\begin{aligned}
    q(\kk,\kp) &= \frac{4\pi^2\xi^4 }{L_x^2L_y^2}
        e^{-\xi^2 (\bm{k}-\bm{k}')^2}
        \left\llangle  \sum_{\alpha,\beta} U_\alpha U_\beta \, \sum_{z,z'} \int\frac{dz_\alpha}{W}  \phi(z-z_\alpha)\phi(z'-z_\beta) \,
        \psi_\kp^\dagger (z) \psi_\kk^\pd (z)
        \psi_\kk^\dagger (z') \psi_\kp^\pd (z') \right\rrangle \\
    &= \frac{4\pi^2\xi^4 \delta^2}{3 L_x^2L_y^2}
        e^{-\xi^2 (\bm{k}-\bm{k}')^2}
        \sum_{\alpha} \, \sum_{z,z'} \,
        \psi_\kp^\dagger (z) \psi_\kk^\pd (z)
        \psi_\kk^\dagger (z') \psi_\kp^\pd (z')\,
        \int\frac{dz_\alpha}{W}  \phi(z-z_\alpha)\phi(z'-z_\alpha) \\
    &=\frac{4\pi^2\xi^4 \delta^2 n_i}{3 L_x L_y}
        e^{-\xi^2 (\bm{k}-\bm{k}')^2}
        \sum_{z,z'} \,
        \psi_\kp^\dagger (z) \psi_\kk^\pd (z)
        \psi_\kk^\dagger (z') \psi_\kp^\pd (z')
        \int dz_i \,  \phi(z-z_i)\phi(z'-z_i) \\
    &= \frac{(2\pi\xi^2)^3 \delta^2 n_i}{3 L_x L_y}
        e^{-\xi^2 (\bm{k}-\bm{k}')^2}
        \sum_{z,z'} \,
        \psi_\kp^\dagger (z) \psi_\kk^\pd (z)
        \psi_\kk^\dagger (z') \psi_\kp^\pd (z')
        M(z,z',\xi,W),
\end{aligned}
\end{equation}
where $n_i = \sum_\alpha/L_xL_yW$ is the impurity
concentration, furthermore we used
\begin{equation}
\left\llangle U_\alpha U_\beta \dots \right\rrangle =
\delta_{\alpha\beta} \frac{\delta^2}{3} \llangle \dots \rrangle,
\end{equation}
and defined the function
\begin{align}
M(z,z',\xi,W) &=
\frac{1}{2\pi \xi^2}\int dz_i \,  \phi(z-z_i)\phi(z'-z_i)
\\
&=\frac{ e^{-(z-z')^2/4\xi^2}}{4\sqrt{\pi}\xi}
    \left[
    \mathrm{erf}\left(
    \frac{W + z+z'}{2\xi}
    \right)
    +
    \mathrm{erf}\left(
    \frac{W - z-z'}{2\xi}
    \right)
    \right],
\end{align}
plotted in Fig.~\ref{fig:mfct}; the error function is defined as
$\mathrm{erf}\left(x\right)=(2/\sqrt{\pi})\int_0^xe^{-t^2}\mathrm{d}t$.
Note that
\begin{equation}
    \int dz'\, M(z,z',\xi,W)
\end{equation}
is a weak function of $z$, equal to $1$ in the middle of the
slab, and going down to $\approx 0.4$ at the edges in the
range $\xi$.

\end{document}